\documentclass[prl,twocolumn,nofootinbib, preprintnumbers, superscriptaddress]{revtex4}

\usepackage{amsmath,amssymb,slashed,braket}
\usepackage{graphicx}
\usepackage{epstopdf}
\usepackage{float}
\usepackage[colorlinks=true,
            linkcolor=blue,
            urlcolor=blue,
            citecolor=green,          
            bookmarks=true,
            bookmarksnumbered=true,
            breaklinks=true,
            pdfpagemode=Fullscreen,
            pdfstartview=FitBH]{hyperref}

\usepackage[normalem]{ulem}
\usepackage{color}

\definecolor{Orange}{cmyk}{0,0.61,0.87,0}
\definecolor{JungleGreen}{cmyk}{0.99,0,0.52,0}
\definecolor{OliveGreen}{cmyk}{0.64,0,0.95,0.40}
\definecolor{Brown}{cmyk}{0,0.81,1,0.60}
\definecolor{RoyalBlue}{cmyk}{0.71,0.53,0,0.12}
\definecolor{Gray}{cmyk}{0,0,0,0.40}
\definecolor{LightPink}{cmyk}{0.0,0.25,0,0}
\definecolor{LLightPink}{cmyk}{0.0,0.10,0,0}
\definecolor{LightBlue}{cmyk}{0.25,0,0,0}
\definecolor{LightGray}{cmyk}{0,0,0,0.2}

\usepackage{xcolor}
\definecolor{gesfpurple}{rgb}{0.47,0.19,0.42}

\definecolor{gesflanse}{rgb}{0.00,0.50,0.50}

\definecolor{gesfblue}{rgb}{0.08,0.42,0.76}

\definecolor{gesfred}{rgb}{1,0,0}

\definecolor{gesfwhite}{rgb}{1,1,1}

\definecolor{gesfblack}{rgb}{0,0,0}

\newcommand{\geqn}[1]{Eq.\,\hypersetup{linkcolor=blue}(\ref{#1})\hypersetup{linkcolor=blue}}
\newcommand{\gfig}[1]{{\hypersetup{linkcolor=violet}Fig.\,\ref{#1}\hypersetup{linkcolor=blue}}}
\newcommand{\gtab}[1]{{\hypersetup{linkcolor=gesflanse}Tab.\,\ref{#1}\hypersetup{linkcolor=blue}}}

\graphicspath{{figs/}}

\begin{document}

\title{Thermal Relic Right-Handed Neutrino Dark
Matter}

\author{Jie Sheng}
\email[Corresponding Author: ]{shengjie04@sjtu.edu.cn}
\affiliation{Tsung-Dao Lee Institute \& School of Physics and Astronomy, Shanghai Jiao Tong University, China}
\affiliation{Key Laboratory for Particle Astrophysics and Cosmology (MOE) \& Shanghai Key Laboratory for Particle Physics and Cosmology, Shanghai Jiao Tong University, Shanghai 200240, China}

\author{Yu Cheng}
\email[Corresponding Author: ]{chengyu@sjtu.edu.cn}
\affiliation{Tsung-Dao Lee Institute \& School of Physics and Astronomy, Shanghai Jiao Tong University, China}
\affiliation{Key Laboratory for Particle Astrophysics and Cosmology (MOE) \& Shanghai Key Laboratory for Particle Physics and Cosmology, Shanghai Jiao Tong University, Shanghai 200240, China}

\author{Tsutomu T. Yanagida}
\email{tsutomu.yanagida@sjtu.edu.cn}
\affiliation{Tsung-Dao Lee Institute \& School of Physics and Astronomy, Shanghai Jiao Tong University, China}
\affiliation{Key Laboratory for Particle Astrophysics and Cosmology (MOE) \& Shanghai Key Laboratory for Particle Physics and Cosmology, Shanghai Jiao Tong University, Shanghai 200240, China}

\begin{abstract}
It is known that two heavy Majorana right-handed neutrinos are sufficient to generate the baryon asymmetry in the present universe. Thus, it is interesting to identify the third right-handed neutrino $N$ with the dark matter. We impose a new discrete symmetry $Z_2$ on this dark matter neutrino to stabilize it. However, the $U(1)_{B-L}$ gauge boson $A'$ couples to the right-handed neutrino $N$. If the $B-L$ breaking scale $V_{B-L}$ is sufficiently low, the dark matter neutrino $N$ can be in the thermal bath. We find that the thermal relic $N$ can explain the dark matter abundance for the $B-L$ breaking scale $ V_{B-L} \sim O(10)$\,TeV. After considering all the constraints from the existing experiments, a narrow mass region of the thermal produced right-handed neutrino dark matter $N$ is still surviving.

\end{abstract}

\maketitle 

{\bf Introduction} -- 
The intriguing $U(1)_{B-L}$ extension of the Standard Model (SM) predicts three right-handed neutrinos (RHNs) to cancel the gauge anomalies. 
The presence of the heavy RHNs is
a key point for the natural explanation of the observed small neutrino masses through the seesaw mechanism \cite{Minkowski:1977sc, Yanagida:1979as, Yanagida:1979gs, GellMann:1980vs}, as well as for 
the creation of the baryon asymmetry in the present Universe via the leptogenesis \cite{Fukugita:1986hr, Buchmuller:2005eh}. 
It has been pointed out \cite{Frampton:2002qc, Endoh:2002wm} that two heavy Majorana neutrinos $N_{1,2}$ are sufficient for generating the observed baryon asymmetry 
\footnote{It is pointed out \cite{King:1999mb} that single right-handed neutrino dominance has a natural explanation for the large mixing angle in the neutrino sector.}. 
This fact has raised a new question, '\textit{Why do we have three families?}'
A simple answer to this question is 
that one more RHN (called as $N$) is required to explain the dark matter \cite{Kusenko:2010ik, Ibe:2016yfo}. 

In the previous works, we have considered the non-thermal production of the RHN DM $N$ \cite{Kusenko:2010ik, Ibe:2016yfo}, 
provided that the breaking scale of the $U(1)_{B-L}$ gauge symmetry is extremely high such as a GUT scale. 
Thus, it is very difficult to test such a model 
since the interaction between DM $N$ and the SM particles is extremely weak. 
However, the sufficient baryon asymmetry 
can be produced at the Universe temperature of $\mathcal{O}(10)\,$TeV if the two Majorana masses of $N_{1,2}$ are almost degenerate \cite{Pilaftsis:1997jf, Xing:2006ms}. 
Therefore, it is interesting to consider the $U(1)_{B-L}$ gauge symmetry is broken at a relatively low scale around $\mathcal{O}(10)\,$TeV and the two Majorana masses for $N_{1,2}$ are also of the same scale. 

In this paper, we consider the thermal production of the RHN DM $N$ in the early Universe assuming the $B-L$ breaking scale 
$V_{B-L} \simeq O(10)$\,TeV to explain the observed DM density \footnote{The DM $N$ can be in the thermal bath even when the $B-L$ breaking scale is very high if we introduce a new light scalar boson $\phi$ couples to $N$ \cite{Cheng:2023hzw, Cheng:2023dau}.}. 
After introducing the model, 
we first show that such scenario can successfully
explain the DM density in the present Universe. 
Then, the constraints from all existing experiments, including the 
fixed-target, collider, and DM direct detection experiments, are discussed. 
The combined constraints from the direct detection and collider exclude the mass parameter region 
$m_N \gtrsim 5$\,GeV. 
The direct $N$ production experiments at the fixed targets give us very strong constraints on the lower mass range where $m_N < 1$\,GeV. 
Even though, a narrow parameter region for $m_N \subset (10$\,MeV, $5$\,GeV$)$ still remains. 
Such a parameter space 
can be tested in the near future experiments.

{\bf The RHN DM Model} -- 
In this paper, we consider a simple extension of the SM which is based on the gauge symmetry $SU(3)\times SU(2)\times U(1)_Y \times U(1)_{B-L}$. This simple extension is extremely attractive since we need to add three RHNs to cancel the gauge anomalies. The presence of RHNs is a key point of explaining the small neutrino masses via the seesaw mechanism \cite{Minkowski:1977sc, Yanagida:1979as, Yanagida:1979gs, GellMann:1980vs, Wilczek:1979hh} and of generating the observed baryon-number asymmetry in the present universe through 
the leptogenesis \cite{Fukugita:1986hr, Buchmuller:2005eh}.

A new Higgs field $\Phi$ is introduced to break the gauged $B-L$ symmetry spontaneously,
and its couplings to RHNs $N_i (i=1-3)$ generate large Majorana masses for them through 
$\frac{1}{2} h_i \Phi N_i N_i$.
The related Lagrangian is, 
\begin{align}
\mathcal{L}
&=\frac{i}{2} \bar{N}_i \gamma^\mu \partial_\mu N_i
+
\left(\lambda_{i \alpha} \bar{N}_i L_\alpha H -\frac{1}{2} M_{R i} \bar{N}_i^c N_i+\text { h.c. }\right)
\nonumber \\
& - 
\frac{1}{2} g_{B-L} \bar{N}_i \gamma^\mu \gamma_5 N_i A_\mu^{\prime}+
g_{B-L} Q_{B-L} \bar{f} \gamma^\mu f A_\mu^{\prime}.
\end{align}
Here, $L_\alpha$, $f$, and $H$
are the SM left-handed lepton
doublets, fermions, and Higgs boson. The index $i$ stands 
for the generation of RHN and 
$\alpha$ for all the species of leptons. 
The vacuum expectation value of $\Phi$ is the breaking scale of 
$B-L$ gauge symmetry $\langle \Phi \rangle = V_{B-L}$. Thus, the RHN mass is $M_{Ri} = h_i V_{B-L}$.
The $B-L$ quantum numbers 
$Q_{B-L}$ in the present model for all particles are shown in \gtab{tab:QBL}.  
The $q_\alpha$, $u_R$, $d_R$ and $e_R$ are the left-handed quark doublets, right-handed up- and down-type quarks, and the right-handed charged leptons, respectively. 
%

\begin{table}[h]
\centering
\caption{$B-L$ charge for different species}
\label{tab:species}
\begin{tabular}{|c@{\hspace{8pt}}|c@{\hspace{8pt}}|c@{\hspace{8pt}}|c@{\hspace{8pt}}|c@{\hspace{8pt}}|c@{\hspace{8pt}}|c@{\hspace{8pt}}|c@{\hspace{8pt}}|c@{\hspace{8pt}}|}
\hline
{\rm Species} & \centering $q_\alpha$ & $u_R$ & $d_R$ & $L_\alpha$ & $e_R$ & $N_i$ & $\Phi$ & $H$ \\
\hline
$Q_{B - L}$ & 1/3 & 1/3 & 1/3 & -1 & -1 & -1 & 2 & 0\\
\hline
\end{tabular}
\label{tab:QBL}
\end{table}

As explained in the introduction,
one of the right-handed neutrinos $N_i (i=1-3)$ is the DM. 
We can choose the third $N_3$ to be this RHN DM without losing the generality and hence $\lambda _{3\alpha}=0$.
To guarantee the stability of DM, we impose a discrete $Z_2$ symmetry. Only the RHN DM $N$ is odd under the $Z_2$ parity while others are even. 
First of all, we exclude the parameter region that, $g_{B-L} \sim \mathcal{O}(1)$. 
In such a region, the right-handed neutrino $N$ is nothing but the so-called WIMP, and it is most likely excluded by the direct detection experiments. 
Therefore, we consider the small gauge coupling $g_{B-L} \ll 1$ region 
and hence the $B-L$ gauge boson $A'$ is relatively light with mass $m_{A'} \ll V_{B-L} = \mathcal{O}(10)\,$TeV. 
On the other hand, we assume the new Higgs boson $\Phi$ has a mass of the order of the $B-L$ breaking scale.
The effects of this Higgs boson at the freeze-out time of our DM $N$
can be neglected. 
Thus, the physics at the decoupling time of the RHN DM is described by only lighter particles, the DM $N$, the $U(1)_{B-L}$ gauge boson $A'$, the SM particles, and their interactions.

{\bf Thermal Production} --
In this section, 
we discuss the thermal production of the RHN DM $N$ and calculate its relic abundance. First of all, the $B-L$ gauge boson $A'$ 
can be in the thermal bath with the SM particles through its decay and inverse decay, $A' \leftrightarrow f + \bar f$.
The equilibrium condition requires
$
  \Gamma_A \simeq  g_{B-L}^2 m_{A'}/12 \pi 
  \geq \mathcal{H} (m_{A'})
$
where $\mathcal{H}(T) = 1.66 \sqrt{g_*} T^2/M_{PL}$ is the 
hubble constant and $M_{PL} = 1.22 \times 10^{19}\,$GeV the planck mass.
Since $m_{A'} = g_{B-L}\times V_{B-L}$, this condition reads to
$g_{B-L} \geq 12 \pi V_{B-L}/M_{PL} \simeq 10^{-13}$, which is easily satisfied.

Then, we consider possible thermal processes that make the RHN $N$ in the thermal bath. The dominant processes are dependent of the mass difference $\Delta$ between
the $B-L$ gauge boson $A'$ and the RHN $N$,
defined as $\Delta \equiv (m_{A'} - m_N) / m_N$. 
We shall discuss 
it in the following three cases.

\begin{figure}[!t]
\centering
\includegraphics[width=0.5
 \textwidth]{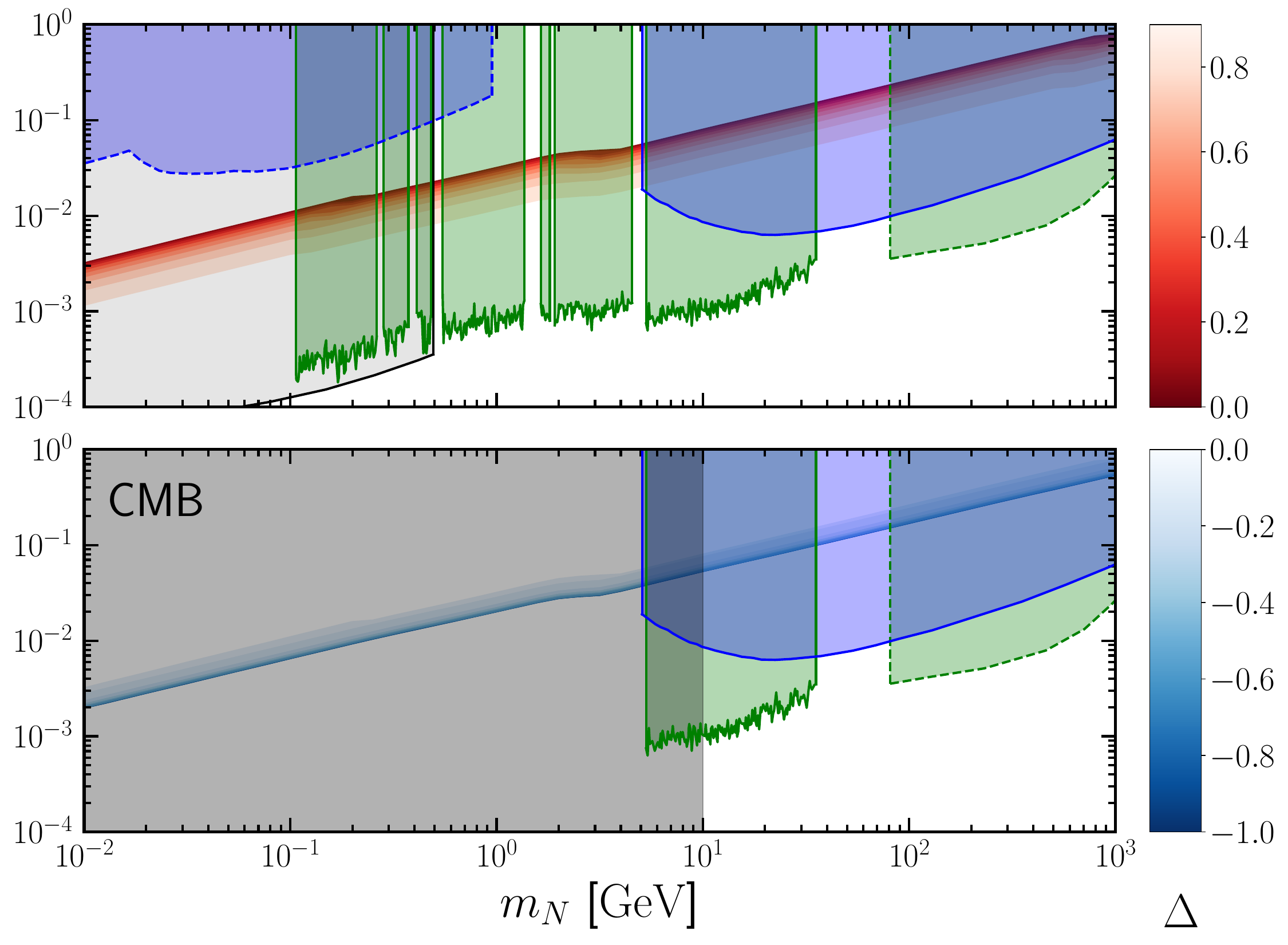}
\caption{
The gradient red (blue) region
in the upper (lower) panel
corresponds to the parameter space which gives the observed DM density for the mass difference $0 < \Delta < 1$ ($\Delta < 0$).
The gray shaded region in the 
lower panel stands for the CMB
constraint while
other colored shaded regions are
excluded by different experiments whose labels 
can be found in FIG.~\ref{fig:RelicDensiy}.
}
\label{fig:RelicANN}
\end{figure}

\noindent \textbf{I. $m_N > m_{A'}$ ($\Delta < 0$):}
In this parameter region, 
there are two processes rendering
the RHN $N$ 
in the thermal bath with the SM particles. First, 
the RHN DM can annihilate into gauge boson
through $N + N\leftrightarrow 2A'$
with a thermal averaged cross section,
\begin{equation}
\langle \sigma v \rangle_{NN \rightarrow 2A'}
= 
\frac{g^4_{B-L}}{4 \pi}
\frac{ 
m_N^2-m_{A^\prime}^2}
{\left(
m_{A^\prime}^2-2 m_N^2
\right)^2}
\sqrt{1-\frac{m_{A^\prime}^2}{m_N^2}}
.
\end{equation}
Second, it also annihilates into SM fermions $f$ and anti-fermion $\bar f$,
$ N + N\leftrightarrow f + {\bar f} $, 
by exchanging the mediator $A'$. The 
thermal averaged cross section,
\begin{equation}
\hspace{-2 mm} \langle \sigma v \rangle_{NN \rightarrow f \bar f}
=
\frac{N_c Q^2_{B-L} g^4_{\rm B-L} ( m_f^2 + 2 m_N^2)
    }{2 x \pi  (m_{A^\prime}^2 - 4 m_N^2)^2}
    \sqrt{1-\frac{m_f^2}{m_N^2}}
\end{equation}
is velocity dependent but with the same order
of $g_{B-L}^4$.

The thermal evolution of 
the RHN yield $Y_N \equiv n_N /s$
(where $n_N$ is the number density 
of RHN and $s = 2 \pi^2 g_* T^3/45$ is the entropy density of the Universe with degrees of freedom $g_*$)
is then described by the Boltzmann 
equation, 
\begin{equation}
    \frac{dY_N}{dx}
    =
    - \frac{s \langle \sigma v \rangle_{\rm ann}}{\mathcal{H} x}
     \left( Y_N^2 - \left( Y_N^{\rm eq} \right)^2 \right),
\label{BoltzmannI}
\end{equation}
where $\langle \sigma v \rangle_{\rm ann}$
is the total annihilation cross section containing both channels.
As the number density decreases, the 
annihilation freezes out. The final yield
can be solved analytically as, 
\begin{equation}
    Y_N^f \approx \frac{3.79 \left(g_*\right)^{-1/2}}{m_N M_{\rm P l} J_f},\quad 
    J_f
    \equiv
    \int_{x_f}^{\infty} d x \frac{\left\langle\sigma v \right\rangle_{\rm ann}}{x^2}.
\end{equation}
Here, $x_f$ is the freeze-out point. The DM relic density is related to $Y^f_N$ as,
\begin{equation}
  \rho_{\chi} = m_N s_0 Y^f_N,
\quad
  \Omega_{N} h^2
=
  \frac{\rho_{N}}{\rho_{c} / h^2},
\label{eq:rho}
\end{equation}
where $s_0 = 2891.2$ cm$^{-3}$ is the entropy density today and 
$\rho_c = 1.05 \times 10^{-5} h^2$\,GeV/cm$^3$ is the critical
density of the Universe.

The parameter space which explains 
the observed DM density today $\Omega_N h^2 = 0.12$ is shown as the gradient blue region 
of the lower panel in \gfig{fig:RelicANN}. 
The mass range $m_N <10$\,GeV has a very serious constraint from the CMB observations \cite{Kawasaki:2021etm,Cline:2013fm}. 
Furthermore, other parameter region 
is also excluded by the existing experimental data. 
More details shall be discussed in the next section.

\noindent \textbf{II. $2 m_N > m_{A'} > m_{N}$ ($0 < \Delta < 1$):} The annihilation of RHN $N$ into gauge 
boson $A'$ is kinetically suppressed
because the gauge boson becomes heavier.
Thus, such a channel is neglected in 
\geqn{BoltzmannI}, and the freeze-out
is determined only by the annihilation 
channel $N + N \leftrightarrow f + \bar f$. 
As shown in the upper panel of 
\gfig{fig:RelicANN},
since the total annihilation
cross section 
does not change too much, the parameter
space (gradient red region) giving the observed DM relic density
is similar to the case that $\Delta < 0$. 
Although there is no CMB constraint
here for the annihilation channel to 
gauge boson is forbidden, 
all the parameter space is still excluded
by experimental constraints.

\noindent \textbf{III. $ m_{A'} > 2 m_{N}$ ($\Delta > 1$):}
Once the gauge boson mass is larger 
than two times of the RHN mass, the 
decay and inverse decay channel\footnote{
A similar model was considered in \cite{Kaneta:2016vkq}. 
However, the most important process, the decay-inverse decay channel, is not taken into account in their work. Hence, most of their parameter spaces for the thermal relic $N$ DM are excluded at present, as we show in FIG.1 in this paper. } 
$A' \leftrightarrow 2 N$ dominants 
with a decay rate 
\begin{equation}
    \Gamma_{A' \rightarrow 2 N} =
    \frac{ g^2_{B-L} m_{A^\prime}}{24 \pi}
    \left( 1 - \frac{4 m_N^2}{m_{A^\prime}^2}\right)^{\frac{3}{2}}
\end{equation}
of order $g_{B-L}^2$. The Boltzmann 
equation in the early Universe 
becomes, 
\begin{equation}
    \frac{d Y_N}{d x} =
    - \frac{1}{\mathcal{H} x}
    \left[ 
    -\Gamma_{A^\prime \rightarrow N N} \left(
    Y_{A^\prime} - \frac{Y_N^2 Y^{\rm eq}_{A^\prime}}{\left(Y_N^{\rm eq}\right)^2}
    \right)
    \right].
\end{equation}
The gauge boson $A'$ keeps maintaining in 
equilibrium for its decay rate is always
larger than the Hubble constant in the 
concerning energy scale. The freeze-out
point $x_f$ can be obtained by the condition that the coefficient of the 
collision
term becomes order of one, 
\begin{equation}
     \Gamma_{A^\prime \rightarrow N N} 
     \frac{ Y^{\rm eq}_{A^\prime}}{Y_N^{\rm eq}} 
     \simeq
\mathcal H x_f.
\label{decayFreeze}
\end{equation}
To match the observed DM density, the 
RHN mass has a relationship with the 
coupling $g_{B-L}$ and mass difference
$\Delta$ as \cite{Frumkin:2021zng}, 
\begin{equation}
    m_N \simeq 
    \left[
    \frac{g^2_{B-L}}{16\pi}
    \left(
    \frac{M_{PL}}{1.66 \sqrt{g_*}}
    \right)
    \left( 
    \frac{2^{3/2} \pi^{7/2} g_* T_{\rm eq}}{60 x_f^{1/2 - 3/\Delta}}
    \right)
    \right]^{\frac{1}{\Delta+1}}.
\end{equation}
Here, we applied the DM number
density, $n_N (x_f) \simeq 
\pi^2 g_* T_{\rm eq} m_N^2/ 30 x_f^2$.

\begin{figure}[!t]
\centering
 \includegraphics[width=0.5
 \textwidth]{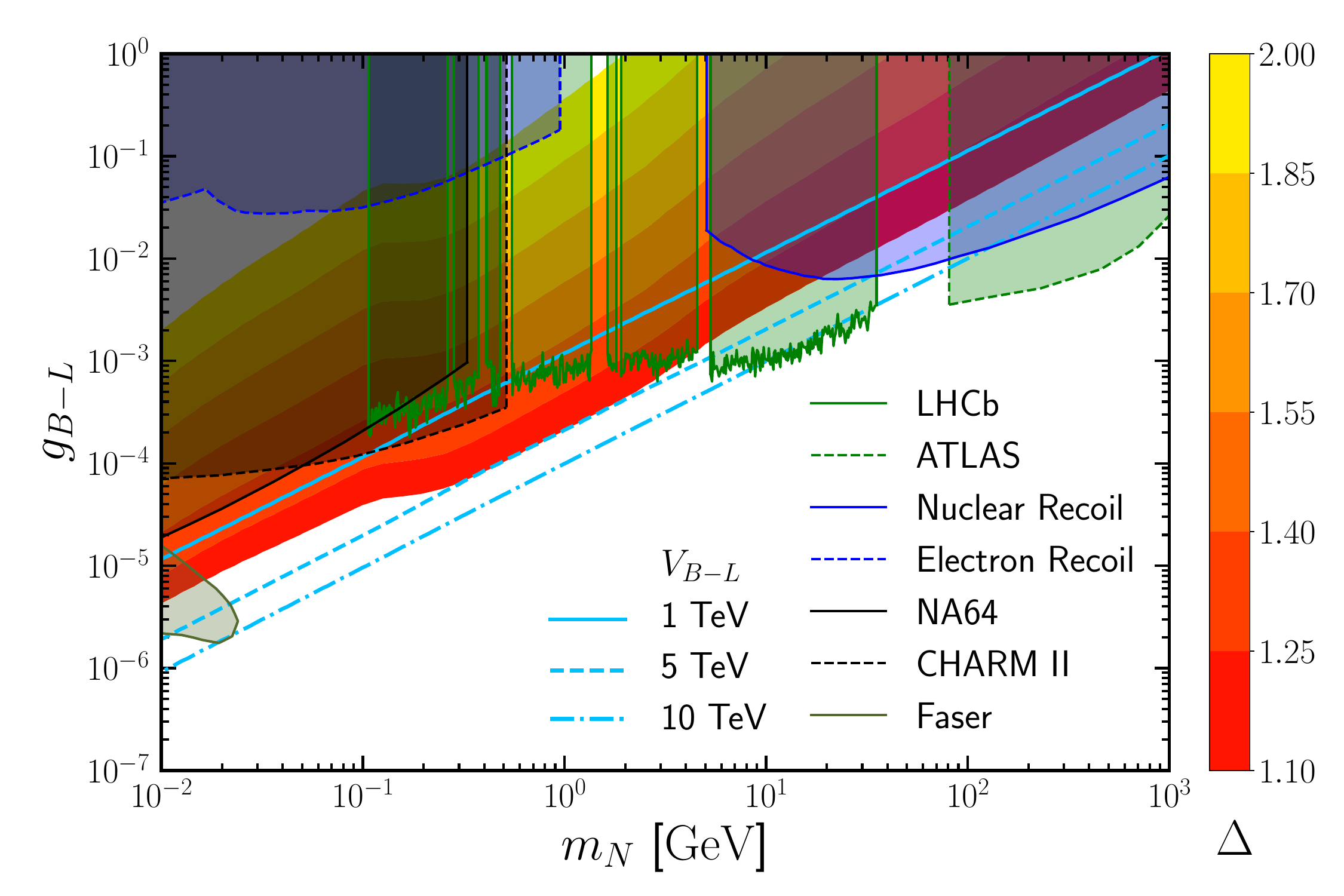}
\caption{
The parameter space (red padding region) which gives the observed DM density for the mass difference $\Delta > 1$.
The colored (green, blue, black) shaded regions are
excluded by different experiments.
The blue lines (solid, dashed, dash-dotted) from above to below correspond to $V_{\rm B-L} = 1,5,10\,$TeV.
}
\label{fig:RelicDensiy}
\end{figure}

The corresponding parameter space 
is shown in \gfig{fig:RelicDensiy} as the 
gradient red padding area. 
The same as the 
standard freeze-out scenario,
a larger DM mass requires a 
larger coupling.
Besides, 
for a fixed DM mass, the $B-L$ gauge boson mass
$m_{A'}$ increases as 
the mass difference $\Delta$ increase.
As a result, the number density of the gauge
boson becomes smaller and a larger 
coupling is needed to compensate for
the decrease of the left-handed side
of \geqn{decayFreeze}. 
One can see
that the DM with a mass 
$m_N \gtrsim 5$\,GeV is excluded by 
DM direct detection and collider 
experiments. However, there is 
still some regions surviving for 
$m_N \subset (10\,$MeV, $5\,$GeV$)$
with the vacuum-expectation value $V_{B-L} \subset (1\,$TeV, $5\,$TeV$)$.

{\bf Experimental Constraints} --
The RHN DM $N$, as well
as a new massive gauge boson $A'$, is
predicted in our model. Both of
them have interactions with 
SM particles and receive strong
constraints from experiments. In this
section, we shall analyze the 
results from fixed-target, collider, 
and DM direct detection experiments
and show the surviving parameter 
space. All the constraints are shown in \gfig{fig:RelicDensiy}.

\noindent \textbf{I. Fixed-Target Experiments:}
The $B-L$ gauge boson $A^\prime$ can be produced by the 
bremsstrahlung process in high-energy electron scattering 
off heavy nucleus, $e^- + Z \rightarrow e^- + Z + A^\prime$. 
Its subsequent invisible decay into a neutrino pair will 
cause missing energy events in fixed-target experiments \cite{Gninenko:2016kpg,LDMX:2018cma,Chen:2022liu}. With no signal events 
found, NA64 experiment data \cite{NA64:2022yly} gives the stringent constraints on the $B-L$ gauge coupling, $g_{B-L} < (2 \times 10^{-5}, 10^{-4})$
according to the mass range $ m_{A'} \subset(10,100)$\,MeV as shown in the gray shaded region surrounded by solid curve.

The extra $B-L$ gauge symmetry also generates
a non-standard $\nu-e$ interaction, 
which shall lead to
some recoil excess in neutrino detectors. 
Thus, the strongest constraint 
for the mass region $m_{A^\prime} \subset ( 0.1, 1)\,$GeV
comes from $\nu-e$ scattering experiments, CHARM II \cite{CHARM-II:1993phx}, which is shown as the gray shaded region surrounded by dashed curve.

\noindent \textbf{II. Collider Search:}
Several searches for the $B-L$ gauge boson produced in proton-proton collisions at LHC were also carried
out by the ATLAS, CMS, LHCb and FASER collaborations \cite{ATLAS:2017fih,CMS:2016cfx,LHCb:2017trq,FASER:2023tle}.
By searching for the new high-mass 
resonances phenomena in dielectron
and dimuon final states, the ATLAS collaboration gives the constraint for $g_{B-L} < 10^{-3}$ in $ m_{A'} \subset(100,1000)$\,GeV \cite{ATLAS:2017fih}
as shown by the green shaded region surrounded by dashed curve. 
An inclusive search for the gauge boson decay $A^\prime \rightarrow \mu^+ + \mu^-$ was performed by LHCb experiment\cite{LHCb:2017trq,Amrith:2018yfb}. By searching for the displaced-vertex
signature for long-lived $A^\prime$ in $2 m_\mu < m_{A^\prime} < 350$\,MeV and 
prompt-like $A^\prime$ decay in $350$\,MeV $< m_{A'} < 70$\,GeV, the experiment constrains $g_{B-L} < 10^{-4}$ as shown by the the green shaded region surrounded by solid curve. 
The FASER experiment is designed to search for long-lived particles travelling in the far-forward direction of the proton beam. It excludes the parameter space where $g_{B-L} \subset (5 \times 10^{-6}, 2 \times 10^{-5})$ and $m_{A'} \subset (15, 40)$\,MeV \cite{FASER:2023tle} as shown in the dark green shaded region surrounded by solid curve.

\noindent \textbf{III. Direct Detection:}
The RHN itself can scatter 
with SM fermions by exchanging a t-channel $B-L$ gauge boson. 
Thus, it is also constrained from the DM direct detection experiments. The scattering is $p$-wave since the RHN DM $N$ is Majorana type.
Taking a general fermion target 
with mass $m_f$, the leading order of the $p$-wave
scattering cross section $\sigma_S$ is, 
\begin{equation}
    \sigma_{S} = \frac{m_N^2 m_f T_N (3 m_N^2 + 2 m_N m_f + m_f^2)}{(m_N + m_f)^4 \pi}.
\end{equation}
Here, $T_N$ is the kinetic energy of initial DM. For the halo DM with
velocity $v \sim 10^{-3}$, the kinetic energy is of order $T_N \sim 10^{-6} m_N$. In the non-relativistic limit, the interaction between Majorana 
RHN and SM fermion via a vector mediator contains both spin-independent 
and spin-dependent contributions\cite{Fan:2010gt}. The spin-independent 
part plays a leading role and receives stronger constraints.

For the RHN DM with mass over $\mathcal{O}(1)$\,GeV, the most sensitive 
constraints are from the Xenon\cite{XENON:2023cxc}, LZ\cite{LZ:2022lsv}, and PandaX\cite{PandaX-4T:2021bab} which are based on the DM-nuclear scattering. For the coherent scattering of Xenon nuclear, the 
target mass is $m_f = 131\,$GeV. As shown in the blue shaded region
surrounded by solid lines, the nuclear recoil constraint is slightly 
weaker than collider ones. However, it fills 
the gap around $(30, 70)\,$GeV so that the possibility of $m_N \gtrsim 5\,$GeV is totally excluded.
On the other hand, one needs to consider the DM-electron scattering ($m_f = m_e$) for the RHN DM with mass $m_N \lesssim \mathcal{O}(1)$\,GeV because of the kinetic threshold.
In this region, the most stringent constraint 
(blue shaded region surrounded by dashed curve) is 
from SENSEI\cite{SENSEI:2020dpa}, DAMIC\cite{DAMIC-M:2023gxo} and Darkside\cite{DarkSide:2022knj}.
It is much weaker than that of fixed-target experiments.

Combining all the constraints, one can clearly see that
the parameter space for $m_{A'} < 2 m_N$ ($\Delta < 1$), which prefers $g_{B-L} > 10^{-3}$, 
has already been excluded. The only remaining region 
is $m_N \subset (10\,$MeV, $5\,$GeV$)$ for 
$\Delta > 1$ and $g_{B-L} \lesssim 10^{-4}$ except for a small region with $m_N < 20$\,MeV which is excluded by the FASER experiment.

{\bf Discussion and Conclusions} --
We have considered, in this paper, the B-L extension of the standard model,  which requires three RHNs to cancel the gauge anomalies.
The two of them should be used to produced lepton asymmetry that is converted to the baryon asymmetry in the president universe. Therefore, 
the third RHN is a 
natural candidate for the dark matter. In this paper we have  explored this interesting possibility postulating a discrete symmetry $Z_2$ to stabilize the third RHN.


It has been pointed out \cite{Pilaftsis:1997jf, Xing:2006ms}
that we can have a successful low-energy scale leptogenesis if the two Majorana RHNs $N_{1,2}$ have almost degenerate masses $M_1 \simeq M_2$. However, there are several dangerous processes which dilutes the produced lepton asymmetry \footnote{Introduction of $\Phi$ with 
mass $m_\Phi \sim M_1 \sim V_{B-L}$ brings the scattering process
$N N \rightarrow N N (H^\dagger H)$ in the thermal bath.}. In order to show the resonant leptogenesis works we adopt here the non-thermal leptogenesis \cite{Buchmuller:2005eh}. We assume the inflaton decay is sufficiently slow and  the reheating temperature is 
$T_R \sim 140\,$GeV without suppressing 
the sphaleron process \cite{Klinkhamer:1984di,Hong:2023zrf}. We see that all dangerous processes 
are smaller 
than the Hubble rate $H \sim 
T_R^2/ M_{PL}$ once $V_{B-L} \gtrsim 3\,$TeV due to the Boltzmann suppression, and never comes 
into equilibrium \cite{Buchmuller:2005eh}. 
Clearly, the most dangerous inverse decay process of $ l + H\rightarrow N_{1,2}$ is also suppressed.

Once the breaking scale is fixed, all the coupling strengths in our scenario are well-predicted and receive strong constraints, which is different from other DM models with intermediate vector boson that have arbitrary couplings \cite{Pospelov:2007mp,Alves:2015pea,Arcadi:2024ukq}.
Motivated by this observation, we consider thermal production of the RHN DM $N$ in this paper and find the allowed parameter space, $m_N \subset (10$\,MeV $,5$\,GeV$)$ and $g_{B-L}  \simeq 10^{-4}$, consistent with the present DM density and with all existing experiments \footnote{The parameter region of $m_N < 10$\,MeV has a tension with the success of BBN.}. 
We show that the higher mass region for $m_N \gtrsim 5$\,GeV is already excluded by the DM direct detection and the LHC experiments. 
The fixed-target experiments give us stringent constraints for the lighter mass region, $m_N < \mathcal{O}(1)$\,GeV. 
Therefore, we expect that the high-luminosity LHC experiments will provide us crucial tests of the present scenario for the RHN DM 
of the mass in the surviving range and/or the $B-L$ gauge boson $A'$ will be discovered in future high intensity electron beam frontiers such as LDMX-style missing momentum experiments \cite{Gninenko:2016kpg,LDMX:2018cma,Chen:2022liu} and Belle II if the mass is in fact $m_{A'} < \mathcal{O}(1)$\,GeV.

Note that the Ref.\cite{Smith:2024jve} 
shows our remaining parameter 
space has also been excluded by  
the constraint from the invisible decay $B \rightarrow K + A'$. 
However, the definition of the $B-L$ charge 
has an ambiguity due to the hypercharge gauge rotation. For 
$Q_{B-L} \rightarrow Q_{B-L} - 2Y$, the charge 
for $Q_{q_\alpha} = 0$. 
Then, the above decay channel is sufficiently suppressed even after the electroweak symmetry breaking, but the  $B-L$ charge of the right-handed neutrino $N$ is unchanged. Therefore, the constraints in \cite{Smith:2024jve} are not applicable in our model with the new $B-L$ charge and our conclusion in this paper remains the same.

\section*{Acknowledgements}
The authors thank Liang Tan for useful discussions. 
This work is supported by 
the National Natural Science
Foundation of China (12175134, 12375101, 12090060, 12090064, and 12247141),
JSPS Grant-in-Aid for Scientific Research
Grants No.\,24H02244, 
the SJTU Double First Class start-up fund No.\,WF220442604,
and World Premier International Research Center
Initiative (WPI Initiative), MEXT, Japan.
T. T. Y.
is an affiliated member of Kavli IPMU, University of Tokyo.

\providecommand{\href}[2]{#2}\begingroup\raggedright\endgroup

\vspace{15mm}
\end{document}